\documentclass[aps,prd,twocolumn,showpacs,nofootinbib,amsmath,amssymb,floatfix,superscriptaddress,showkeys]{revtex4-1}
\usepackage{graphicx}
\usepackage{dcolumn}
\usepackage{bm}
\usepackage{captcont}
\usepackage{xcolor}
\usepackage[colorlinks,allcolors=blue]{hyperref}
\usepackage{supertabular}

\usepackage{epstopdf}
\usepackage{mathtools}
\usepackage{natbib}
\usepackage{ulem}
\usepackage{hyperref}

\newcommand{\jcap}{J. Cosmol. Astropart. Phys.}

\newcommand{\mnras}{Mon. Not. R. Astron. Soc.}

\newcommand{\aap}{Astron. Astrophys}

\newcommand{\pasa}{Pub. Astron. Soc. Aust.}

\begin{document}

\title{Constraints on primordial black holes in dSphs using radio observations}

\author{Tian-Ci Liu}
\affiliation{Guangxi Key Laboratory for Relativistic Astrophysics, School of Physical Science and Technology, Guangxi University, Nanning 530004, China}
\author{Xiao-Song Hu}
\affiliation{Guangxi Key Laboratory for Relativistic Astrophysics, School of Physical Science and Technology, Guangxi University, Nanning 530004, China}
\author{Yun-Feng Liang}
\email[]{liangyf@gxu.edu.cn}
\affiliation{Guangxi Key Laboratory for Relativistic Astrophysics, School of Physical Science and Technology, Guangxi University, Nanning 530004, China}
\author{Ben-Yang Zhu}
\affiliation{Key Laboratory of Dark Matter and Space Astronomy,
Purple Mountain Observatory, Chinese Academy of Sciences, Nanjing 210023, China}
\author{Xing-Fu Zhang}
\affiliation{School of Astronomy and Space Science, Nanjing University, Nanjing 210023, China}
\affiliation{Key laboratory of Modern Astronomy and Astrophysics (Nanjing University), Ministry of Education, Nanjing 210023, China}
\author{En-Wei Liang}
\email[]{lew@gxu.edu.cn}
\affiliation{Guangxi Key Laboratory for Relativistic Astrophysics, School of Physical Science and Technology, Guangxi University, Nanning 530004, China}
\date{\today}

\begin{abstract}
Primordial black holes (PBHs) are hypothetical objects formed at the early epoch of the universe, which could be a type of dark matter (DM) candidate without the need for new particles. The abundance of PBH DM has been constrained strictly by many observations.
In this work, with the radio observations of Fornax and Segue I, we constrain the abundance of PBH in dwarf spheroidal galaxies through the synchrotron self-Compton (SSC) effect of Hawking radiation electrons.
By selecting optimal sources, we obtain the constraints on the fraction of PBH DM down to {$\sim10^{-3}$ for Segue I and $\sim10^{-5}$} for Fornax at asteroidal mass. 
We also predict that, with 100 hours of future observation by the Square Kilometer Array, the SSC approach could place constraints comparable to the current strictest results for PBHs of $<5\times10^{15}\,{\rm g}$. Better projected constraints can be obtained by including the inverse Compton scattering on cosmic microwave background photons.
\end{abstract}



\maketitle

\section{Introduction}
Primordial black holes (PBHs) have been studied for half a century since Stephen Hawking predicted this kind of black holes with extremely low mass theoretically \cite{1971MNRAS.152...75H}. 
PBHs could also be a type of candidate to explain the measured dark matter (DM) abundance of our universe \cite{1975Natur.253..251C}.
Although no evidence has been found to prove the existence of PBHs, in recent years there has been a resurgence of the research that dark matter is composed of PBH \cite{PhysRevLett.117.061101,2016PhRvL.116x1102A} because of the discovery of gravitational waves (GWs) by LIGO/VIRGO \cite{PhysRevLett.116.061102}. 

PBHs formed in the very early universe. An overdense region from  $\delta\sim$1 fluctuation has a mass larger than the Jeans mass and could collapse to a black hole when it reenters the horizon. 
For such a special formation process, the PBHs have a very wide mass range even down to the Planck mass \cite{2010PhRvD..81j4019C}. 
Therefore, Hawking evaporation could have an obvious effect since the evaporation rate and temperature are inversely related to the PBH mass \cite{1974Natur.248...30H,1975CMaPh..43..199H}.

The abundance of low mass PBHs in the whole universe and in the Galactic Center region has been constrained stringently with the Hawking radiation by many research works \cite{2020PhRvD.101b3010A,2021PhRvD.103j3025I,2020MNRAS.497.1212C,2020PhRvL.125j1101D,2022PhRvD.106l3508Y}, but the PBH abundance in dwarf spheroidal galaxies (dSphs) has not been studied/constrained too much. 
DSphs are promising targets for DM searches since their high mass-to-light ratio suggests they are DM-dominated objects. 
For PBHs with a lifetime larger than the age of our universe, the Hawking radiation is emitted at energies of $<100$ MeV such that the Fermi satellite (which is usually used to detect weakly interacting massive particles) cannot detect them.
Current radio observations of dSphs also cannot give an effective constraint by either synchrotron (SYN) radiation or inverse Compton (IC) scattering of Hawking radiation electrons. 
Ref.~\cite{2021JCAP...03..011D} gives a prediction of constraints on PBH abundance in dSphs using the sensitivity of the upcoming radio telescope Square Kilometer Array (SKA) through IC scattering off cosmic microwave background (CMB) photons.

In this work, we attempt to constrain the PBH abundance in nearby dSphs of the Milky Way by considering the synchrotron self-Compton (SSC) effect. The SSC peak frequency is close to the frequency band of the existing radio observations (see Sec.~\ref{sect:method_cal}) for a PBH mass of $\sim10^{15}\,\rm g$.

The outline of this work is as follows: In Section \ref{sect:PBH}, the formation and evaporation of PBHs are briefly introduced.
In Section \ref{sect:method_cal}, the constraining method is described in detail. We also discuss the effect of the propagation equation in flux calculation. 
The results are shown in Section \ref{sect:results} and finally we summarize in Section \ref{sect:conclusion}.

\section{PBH evaporation}
\label{sect:PBH}

PBH is a special kind of Massive Compact Halo Objects (MACHOs). Considering the nature that they interact with normal matter only through gravity and have existed since the early universe, PBH is a suitable DM candidate.

PBHs are generally believed to be formed from the spherical collapse when the fluctuations reenter the Hubble horizon during the radiation-dominated era \cite{2022arXiv220204653A}. Therefore PBH mass is in the same order as the mass enclosed in Hubble horizon ($M_{\rm H}$) with a factor $\gamma\sim0.2$ \cite{2016PhRvD..94h3504C,2021RPPh...84k6902C},
\begin{equation}
\begin{aligned}
M_{\rm PBH} &= \gamma M_{\rm H}\mid_{\rm form} \sim 10^{15}\gamma\left(\frac{t}{10^{-23}\rm s}\right)\,{\rm g}.
\end{aligned}
\end{equation}

There are various PBH mass functions inferred from different forming theories \cite{1975ApJ...201....1C,1998PhRvL..80.5481N, PhysRevD.47.4244}, where the monochromatic form is the most representative. Although an exact monochromatic mass function is not realistic physically and the mass function is generally expected to be extended in an inflationary scenario, the monochromatic limits can be converted to apply for various extended mass distributions of PBHs \cite{2010PhRvD..81j4019C,1975ApJ...201....1C}. In this work, we only adopt the monochromatic PBH distribution.

The PBH mass is proportional to the time from the Big Bang since the energy density is $\rho\varpropto t^{-2}$ in the early universe.
Therefore, PBHs forming from the Plank time ($10^{-43}\,{\rm s}$) to $1\,{\rm s}$ have a very wide mass range from Plank mass ($10^{-5}$g) to $10^5 M_{\odot}$ \cite{2010PhRvD..81j4019C}. On the other hand, PBHs of low mass have an obvious effect according to Hawking's theory that black holes could emit all types of standard model particles with the temperature \cite{1974Natur.248...30H,1975CMaPh..43..199H}
\begin{equation}
T_{\rm BH} = \frac{\hbar c^3}{8\pi k_{\rm B} GM} \sim 10^{-7}\left(\frac{M_\odot}{M}\right){\rm K}
\end{equation}
with $k_{\rm B}$ and $G$ the Boltzmann and gravitational constants, respectively.

Hawking radiation includes primary and secondary components. The primary emission is a graybody spectrum expressed as \cite{2010PhRvD..81j4019C,PhysRevD.41.3052}
\begin{equation}
\frac{{\rm d}N_e}{{\rm d}E{\rm d}t} = \frac{1}{2\pi\hbar}\frac{\Gamma_e}{\exp\left({{E}/{k_{\rm B}T_{\rm BH}}}\right)-{(-1)}^{2s}},
\end{equation}
where $s$ is the spin of the particle and $\Gamma_e$ is the absorption coefficient.
The secondary emission is produced by fragmentation and decay of primary emission particles and is dominated by $2\gamma$ decay of neutral pions.

As a result of Hawking radiation, the lifetime of evaporating PBHs is \cite{2010PhRvD..81j4019C}
\begin{equation}
\tau \sim \frac{G^2M^3}{\hbar c^4} \sim 10^{64}\left(\frac{M}{M_\odot}\right)^3{\rm yr}.
\end{equation}
It indicates that the PBH below $M_*\sim(5\times10^{14}\rm g)$ has been totally evaporated with a Planck relic left. Due to the stability of DM, we choose $M_*$ to be the lower mass boundary in this work.

\section{Method and calculation}
\label{sect:method_cal}

\subsection{Method}
\label{sect:method}
In this work, we consider the radio observations from the Murchison Widefield Array (MWA) and the Giant Metrewave Radio Telescope (GMRT).
The MWA is a low-frequency SKA precursor located in Western Australia, where the Galactic and Extragalactic All-sky MWA (GLEAM) is a survey of the entire radio sky south of declination $+25^{\circ}$ at frequencies between $72\,{\rm MHz}$ - $231\,{\rm MHz}$ \cite{2015PASA...32...25W,2017MNRAS.464.1146H}.
For GMRT, a survey of the 150 MHz radio sky has been released as part of the TIFR GMRT Sky Survey (TGSS) project \cite{2017A&A...598A..78I}.
A combination of the GLEAM survey and the TGSS ADR1 is used in \cite{2019PhRvD.100d3002K} to derive the upper limit surface brightness of 14 dwarf spheroidal galaxies.

A typical low mass PBH (e.g., $10^{15}\,{\rm g}$) has a peak frequency of synchrotron radiation of $\sim10\,{\rm kHz}$, which is too low compared with the MWA/GMRT waveband, and therefore can not be constrained by the MWA or GMRT observation. On the other hand, the peak frequency of the IC process on CMB is about $100\,{\rm THz}$, which is conversely too high to be constrained by the MWA/GMRT. 
The SKA in the first phase is planned to be sensitive in the frequency range up to $50\,\rm GHz$ \cite{2019PhRvD..99b1302K,2021JCAP...09..025C}, with the frequencies closer to the peak frequency of the IC/CMB process of PBHs.
There have been studies predicting the sensitivity of using SKA data to constrain the PBH abundance in dSphs \cite{2021JCAP...03..011D}.
However, a realistic constraint with current existing radio observation is still absent. We note that the peak frequency of the SSC emission of Hawking radiation electrons is $\sim10\,\rm MHz$ and close to the MWA/GMRT waveband so that we could use the MWA/GMRT observations to place constraints on the PBH abundance.

\begin{figure}[b]
\begin{center}
\includegraphics[angle=0,width=0.5\textwidth]{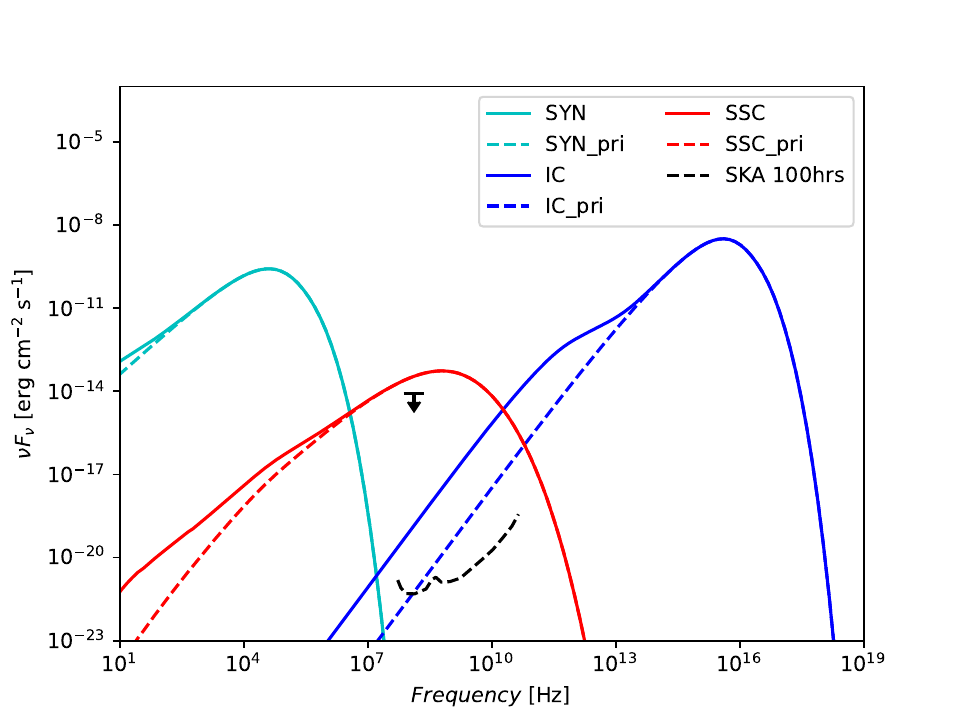}
\end{center}
\caption{The energy spectra for the synchrotron radiation (cyan), IC/CMB process (blue), and SSC effect (red) of $10^{15}\,\rm g$ PBHs in the dSph Segue I. The PBH abundance is assumed to be $1$. The SSC effect is in fact less efficient and has a peak flux several orders of magnitude lower than the synchrotron and IC/CMB processes. But at the frequencies of MWA and SKA, its flux is higher. The solid lines and dashed lines represent the spectra considering and not considering the secondary emission of Hawking evaporation, respectively. 
The black arrow indicates the upper limit on the radio flux obtained from combining MWA and GMRT observations \cite{2019PhRvD.100d3002K}. The SKA sensitivity for 100 hours of observations is shown as a black dashed line.}
\label{fig:spectra}
\end{figure}

In Fig.~\ref{fig:spectra} we show the energy spectra of 3 different processes (SYN, IC and SSC) from Hawking radiation electrons and positrons. The PBH mass chosen for the demonstration is $10^{15}\,\rm g$ with a monochromatic mass spectrum. We consider the dSph Segue I. 
As shown, although the SSC process is actually less efficient, with the peak flux several orders of magnitude lower than the other two mechanisms, its flux is higher than those of the other two processes at the frequencies between 10 MHz and 10 GHz, overlapping with the MWA/GMRT waveband as well as a part of SKA frequency range. 
The fact that the SSC peak flux appears close to the SKA sensitivity range and the MWA/GMRT radio observations indicates that we can place constraints on the PBH based on the existing radio observations, and may make the future results with the SKA more stringent (than those considering only the SYN+IC/CMB processes).

The difference between the fluxes with or without secondary emission is also shown in Fig.~\ref{fig:spectra}. The peak frequency of secondary emission is lower than the primary emission such that the influence of considering the secondary emission is insignificant for the SSC effect since its main contribution is outside the frequency range of the MWA/GMRT and SKA. 
In contrast, for the IC/CMB process, the secondary emission is non-negligible at these frequencies and the constraints will be more stringent considering the secondary emission.

\subsection{Flux calculation}
\label{sect:calculation}

For calculating the SSC flux, the equilibrium density of electrons and positrons produced by PBHs in dSphs is necessary, which can be obtained by solving the propagation equation \cite{2021JCAP...09..025C,2006A&A...455...21C,2020PhRvD.101b3015K,2017arXiv171110989Y}
\begin{equation}
D(E)\nabla^2\left(\frac{{\rm d}n_e}{{\rm d}E}\right)+\frac{\partial}{\partial E}\left[b(E)\frac{{\rm d}n_e}{{\rm d}E}\right]+Q_e=0,
\label{eq:prop}
\end{equation}
where $D(E)$=${D_0}$($E$/{1 GeV})$^{\gamma}$ is the diffusion coefficient. The $b(E)$ is the energy loss due to IC process, synchrotron radiation, Coulomb loss and bremsstrahlung \cite{2006A&A...455...21C},
\begin{equation}
\begin{aligned}
b(E)&=b^0_{\rm IC}\left(\frac{E}{\rm GeV}\right)^2+b^0_{\rm syn}\left(\frac{E}{\rm GeV}\right)^2\left(\frac{B}{\rm \mu G}\right)^2\\&
+b^0_{\rm Coul}n_e\left[1+{{\rm log}\left(\frac{E/m_e}{n_e}\right)}/{75}\right]\\&
+b^0_{\rm brem}n_e\left[{\rm log}\left(\frac{E/m_e}{n_e}\right)+0.36\right]
\end{aligned}
\end{equation}
with $m_e$ the electron mass and $n_e\sim 10^{-6}{\rm cm^{-3}}$ the average number density of thermal electrons in dSphs \cite{10.1093/mnras/stv127,2024PhRvD.109f3036H}.
The values of energy loss coefficients are taken to be $b^0_{\rm IC}\simeq0.25$, $b^0_{\rm syn}\simeq0.0254$, $b^0_{\rm Coul}\simeq6.13$ and $b^0_{\rm brem}\simeq1.51$ in units of $10^{-16}$ GeV/s \cite{Beck_2016,McDaniel_2017}.
The source term for PBH injection is
\begin{equation}
Q_e(E,r) = \frac{f_{\rm PBH}\rho(r)}{M_{\rm PBH}}\frac{{\rm d}N_e}{{\rm d}E{\rm d}t}
\label{eq:inject}
\end{equation}
where $f_{\rm PBH}$ is the fraction of DM in the form of PBHs, and the Hawking radiation spectrum ${{\rm d}N_e}/{{\rm d}E{\rm d}t}$ can be obtained from the code {\tt BlackHawk} \cite{2021EPJC...81..910A,2015CoPhC.191..159S}. We assume that PBHs are Schwarzschild black holes and have a monochromatic mass function. 

Eq.~(\ref{eq:inject}) implicitly assumes that the PBH density is proportional to the total DM density through a factor of $f_{\rm PBH}$. However, there could be some dynamical processes (e.g. mass segregation \cite{2009MNRAS.395.1449A}) leading to a more concentrated PBH distribution. 
In this context, as long as the total number of PBHs remains unchanged, the PBH evaporation will produce the same amount of electrons. Since the CMB photons and the assumed magnetic field are uniformly distributed within dSphs, altering the PBH distribution would not affect the results related to the synchrotron radiation and IC/CMB process. 
However, for the SSC component, a more concentrated distribution of PBHs will produce more centrally concentrated soft photons. 
This would produce a stronger SSC signal, meaning that the results presented in this paper are conservative.
In addition, as for the models with radially dependent magnetic fields \cite{2017JCAP...09..027M, 2021JCAP...09..025C}, a higher central concentration of PBHs may further strengthen the constraints through the SSC effect.

With the boundary condition ${{\rm d}n_e}/{{\rm d}E}(r_h)=0$, the solution of Eq.~(\ref{eq:prop}) can be expressed as \cite{2006A&A...455...21C,2024PhRvD.109f3036H,PhysRevD.101.023015,McDaniel_2017}
\begin{equation} 
\frac{{\rm d}n_e}{{\rm d}E}(r,E)=\frac{1}{b(E)}\int_{E}^{m_{\chi}}{\rm d}E^{\prime}G(r,\Delta v)Q_{e}(E^{\prime},r),
\label{eq:solu}
\end{equation}
where $r_h$ is the diffusion radius.
The Green function $G(r,\Delta v)$ has the form of \cite{2006A&A...455...21C,2024PhRvD.109f3036H,PhysRevD.101.023015,McDaniel_2017}
\begin{equation}
\begin{aligned}
G(r,\Delta v) &=\frac{1}{\sqrt{4\pi\Delta v}}\sum_{n=-\infty}^{n=\infty}(-1)^{n}\int_{0}^{r_{h}}{\rm d}r^{\prime}\frac{r^{\prime}}{r_{n}}\left(\frac{\rho(r^{\prime})}{\rho(r)}\right) \\& \left[\exp\left(-\frac{\left(r^{\prime}-r_{n}\right)^{2}}{4\Delta v}\right)-\exp\left(-\frac{\left(r^{\prime}+r_{n}\right)^{2}}{4\Delta v}\right)\right]
\end{aligned}
\label{eq:green}
\end{equation}
with
\begin{equation}
	r_n=(-1)^nr+2nr_h, \quad \Delta v=\int_{E}^{E'}{\rm d}\tilde{E} \frac{D(\tilde{E})}{b(\tilde{E})}.
	\label{term2}
\end{equation}
We call the solution Eq.~(\ref{eq:solu}) an SD+b scenario (spatial diffusion + energy loss term $b(E)$).

By comparing the time scales of energy loss $\tau_{\rm loss}=E/b(E)$ and diffusion $\tau_{\rm D}\simeq r_h^2/D(E)$ \cite{2006A&A...455...21C}, we can know whether an energy loss or a diffusion process dominates the propagation.
Since the magnetic field is weak, for the source with small diffusion radius $r_h$ (e.g., Segue I \cite{2021JCAP...09..025C}), we have $\tau_{\rm D}\ll \tau_{\rm loss}$.
In this case, we can neglect the energy loss term and the solution of Eq.~(\ref{eq:prop}) is simplified to \cite{2020PhRvD.101b3015K} 
\begin{equation}
\frac{dn_e}{dE}(E,r) = \frac{1}{D(E)}f(r,r_h)Q(E,r).
\label{eq:nb}
\end{equation}
where $f(r,r_h)$ has the form of
\begin{equation}
f(r,r_h)=\int_{r_1=r}^{r_h}{\rm d}r_1\left(\frac{1}{r_1^2}\right)\int_{r_2=0}^{r_1}\frac{\rho(r_2)}{\rho(r)}r_2^2{\rm d}r_2.
\end{equation}
We call this solution (Eq.~\ref{eq:nb}) an Nb scenario (neglecting $b(E)$).

Electron spectra for different PBH masses from $5\times10^{14}\,\rm g$ to $5\times10^{17}\,\rm g$ are calculated with the code {\tt BlackHawk}. 
PBHs with masses below $5\times10^{14}\,\rm g$ cannot be a stable DM candidate since they have been completely evaporated to date, and the Hawking radiation of PBHs with masses above $5\times10^{17}\,\rm g$ is too {weak} to be used to constrain the PBHs with radio data.

\begin{figure}[t]
\includegraphics[angle=0,width=0.47\textwidth]{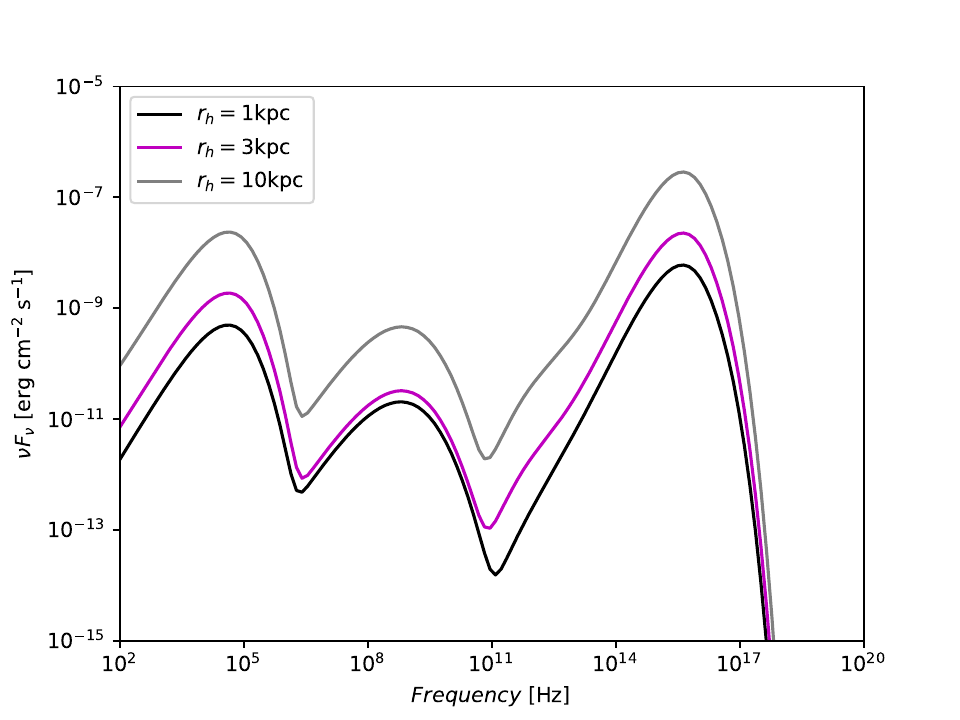}
\caption{The energy spectra of Fornax for different diffusion 
radii $r_h$. A $10^{15}\rm g$ PBH mass and diffusion choice 3 (see Sec.~\ref{sect:results}) are adopted in this plot.}
\label{fig:rh}
\end{figure}

We assume the DM density profile of Fornax to be a Navarro-Frenk-White (NFW) profile \cite{1996ApJ...462..563N,2019PhRvD.100d3002K}
\begin{equation}
\rho_{\rm NFW}(r)=\frac{\rho_s}{(r/r_s)^\gamma(1+r/r_s)^{3-\gamma}},
\end{equation}
and Segue I to be an Einasto profile \cite{1965TrAlm...5...87E,2021JCAP...03..011D}
\begin{equation}
\rho_{\rm EIN}(r)=\rho_s{\rm e}^{-2n[(r/r_s)^{1/n}-1]}.
\end{equation}
The profile parameters of dSphs are taken from Refs.~\cite{2022PhRvD.106l3032D} and \cite{2011JCAP...06..035A} respectively.
To calculate the soft photon field generated from the synchrotron process, the magnetic field strength in dSph is necessary, of which a direct measurement is however still lacking.
There are many theoretical arguments proposed supporting values of $\sim\mu\rm G$ levels \cite{2019PhRvD.100d3002K,2015MNRAS.448.3747R,2011A&A...529A..94C,2017JCAP...07..025R}.
In this work, we adopt a value of $B=1\,\rm \mu G$.
A smaller magnetic field will result in a lower model-expected signal \cite{2021JCAP...09..025C,2024PhRvD.109f3036H} and a weaker constraint on the PBH abundance.

The SSC flux is calculated with the {\tt Naima} package \cite{zabalza2015naima} and we consider the contributions from both electrons and positrons. We let the theoretical SSC flux not exceed the upper $1\sigma$ bound of radio observations \cite{2019PhRvD.100d3002K} or the SKA sensitivity curve to obtain the upper limits on DM abundance for different PBH masses.

We find that the diffusion radius and the distance to the Earth are two critical parameters of dSphs that influence the predicted flux induced by PBH evaporation. 
In this work, we choose Segue I and Fornax to derive the upper limits on PBH abundance because of their short distance (Segue I) and large diffusion radius (Fornax). 
For the diffusion radius, $r_h=2 r_{\rm max}$ is often adopted \cite{Bhattacharjee_2021}, where $r_{\rm max}$ is the distance from the dSph center to the outermost star of the dSph \cite{2015ApJ...801...74G}.
Fig.~\ref{fig:rh} shows the effect of different $r_h$ on the energy spectra of the source Fornax. {For the below results, we also use $r_h=2 r_{\rm max}$ for Fornax, while for source Segue I, we adopt $r_h=1.6$ kpc in \cite{2017JCAP...09..027M}.}
The information/parameters of these two sources in Table \ref{table:info} are taken from \cite{2015PhRvL.115w1301A,2022PhRvD.106l3032D,2011JCAP...06..035A,2015ApJ...801...74G,2017JCAP...09..027M}.


\begin{table*}
	\begin{center}
	\caption{The information and parameters of two sources.}
	\label{table:info}
	\renewcommand\arraystretch{1.5}
	\begin{tabular}{l|ccccccc}
	\hline
	Source & DM profile & index & $\rho_s$ ($M_{\odot}/{\rm kpc}^3$) & $r_s$ (kpc) & Distance (kpc) & $r_{h}$ (kpc) & Refs.\\
        \hline
    Fornax & NFW & $\gamma=1$ & $1.2\times10^7$  & 2.75 & 147 & 12.54 & \cite{2022PhRvD.106l3032D,2015PhRvL.115w1301A,2015ApJ...801...74G}\\
    Segue I & Einasto & $n=3.3$ & $1.1\times10^8$ & 0.15 & 23 & 1.6 & \cite{2011JCAP...06..035A,2017JCAP...09..027M,2015ApJ...801...74G}\\
	\hline
	\end{tabular}
	\end{center}
\end{table*}

\begin{figure}[t]
\begin{center}
\includegraphics[angle=0,width=0.47\textwidth]{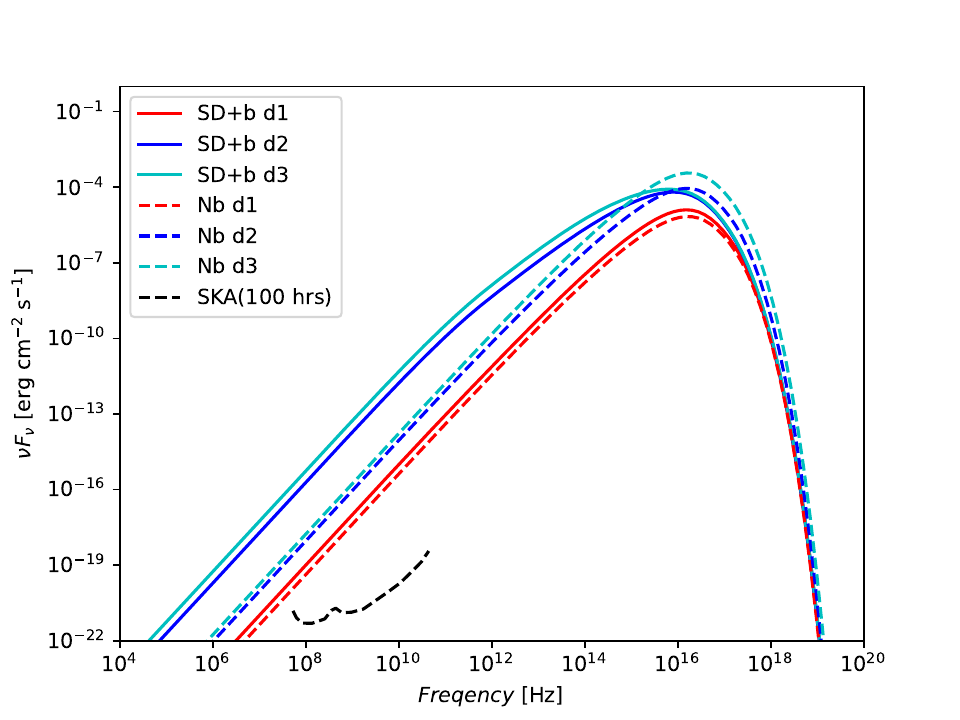}
\end{center}
\caption{The energy spectra for the IC/CMB process for $5\times10^{14}\,\rm g$ PBHs considering only primary emission. The SD+b (solid lines) and Nb (dashed lines) represent the scenarios considering or ignoring the energy loss in calculating the electron propagation. The SKA sensitivity curve for 100 hours of observations is shown as a black dashed line. The energy loss has the effect of shifting the spectra to lower frequencies.}
\label{fig:nb}
\end{figure}

Note that the energy loss rate $b(E)$ is dominated by the synchrotron and IC processes, and it is $b(E)\propto E^2$. Therefore at high energies, $\tau_{\rm loss}$ would be comparable to or smaller than $\tau_{\rm D}$, indicating that the energy loss is predominant and can not be ignored (see Fig.~\ref{fig:nb} for a comparison between the spectra ignoring and not ignoring the energy loss). 
Therefore, for the source Fornax we adopt the SD+b solution.
However, for the source Segue I we find that the quantity $\Delta v$ in Eq.~(\ref{term2}) becomes far greater than the diffusion radius, leading to some numerical issues in the calculation of ${\rm d}n_e/{\rm d}E$.
So we adopt the Nb solution for this source. Considering the smaller Virial radius, we believe this is a good approximation.
Nevertheless, the main results of this work come from the source Fornax.

\begin{figure*}[t]
\begin{center}
\includegraphics[angle=0,width=0.47\textwidth]{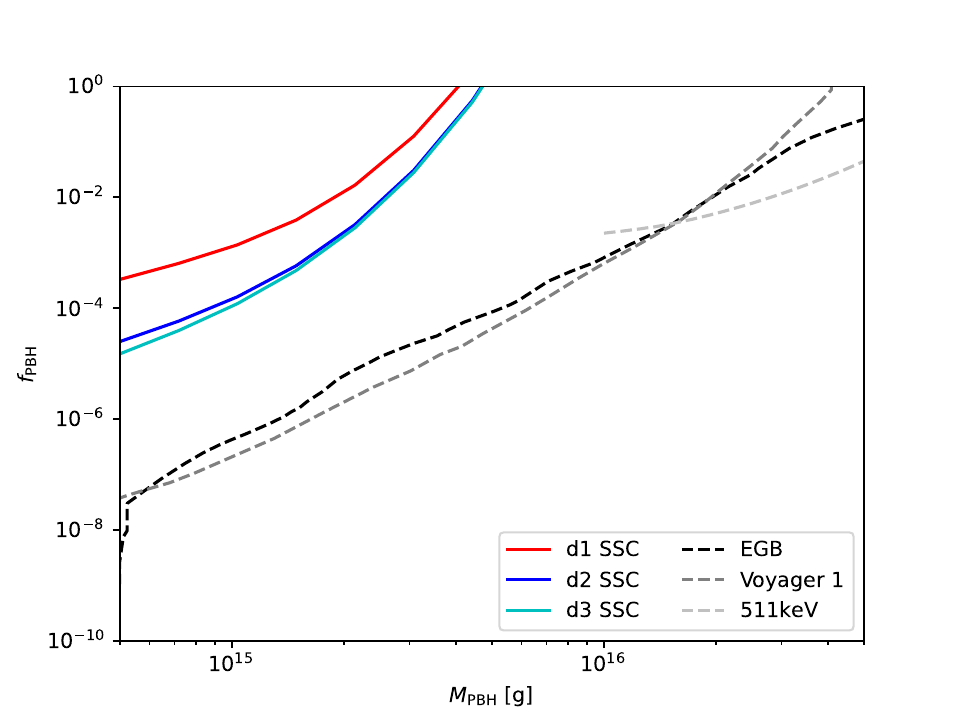}
\includegraphics[angle=0,width=0.47\textwidth]{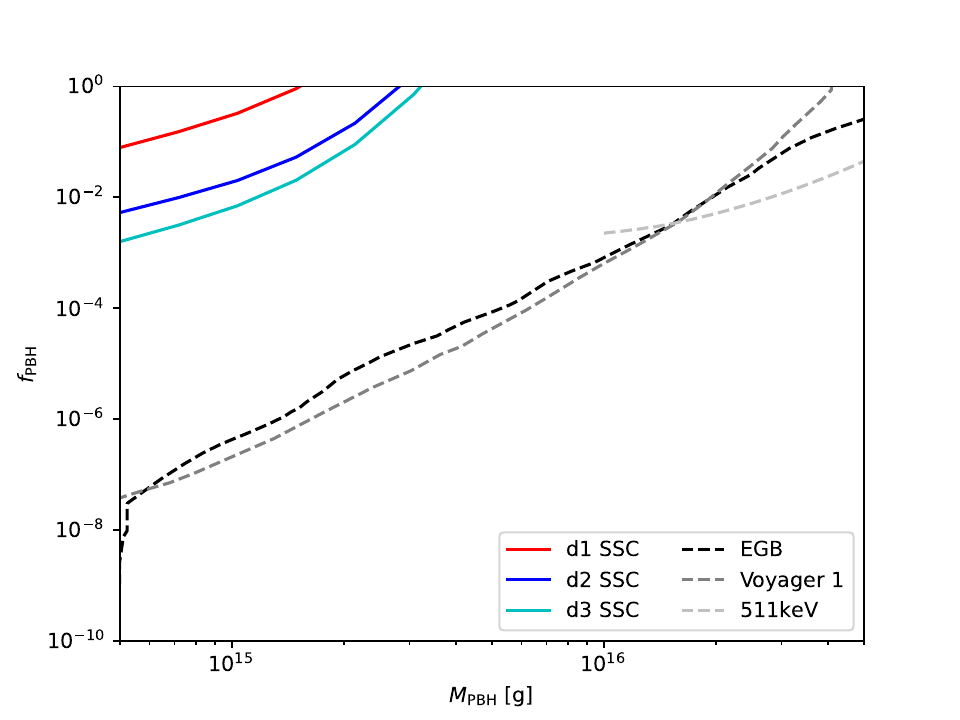}
\end{center}
\caption{Constraints on the PBH abundance in dSphs from the combining MWA and GMRT radio observations of Fornax (left panel) and Segue 1 (right panel) by considering the SSC effect. Different diffusion choices (d1, d2 and d3) are considered. 
Also shown for comparison are results from \cite{PhysRevLett.122.041104,2010PhRvD..81j4019C,PhysRevLett.123.251101} (dashed lines) based on the extragalactic $\gamma$-ray background, Voyager 1 $e^\pm$ spectrum and 511 keV $\gamma$-ray line measurements.
}
\label{fig:mwa}
\end{figure*}

\begin{figure*}[t]
\begin{center}
\includegraphics[angle=0,width=0.6\textwidth]{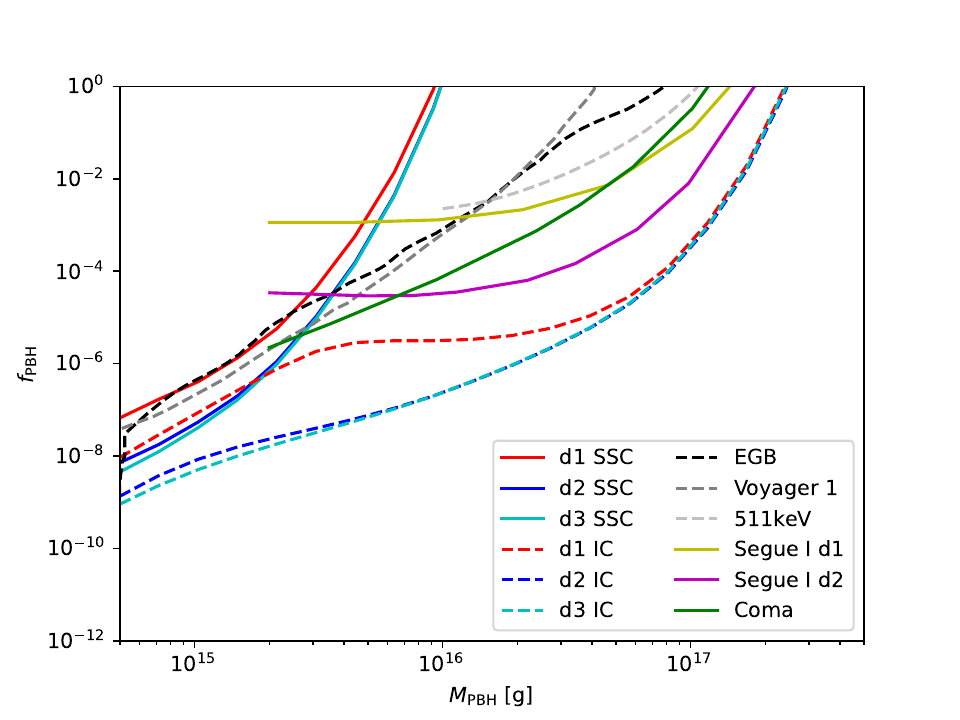}
\end{center}
\caption{The projected constraints on PBH abundance in dSphs for future 100-hour observations of SKA. The three SSC (IC) lines are for the results derived considering the SSC (IC/CMB) flux from PBH Hawking radiation electrons. The results in \cite{2021JCAP...03..011D} which are also predictions for SKA are shown for comparison (the Segue I and Coma lines). Note that the diffusion choices of Segue I (yellow and magenta lines) are the same as diffusion choices 1 and 2 in this work.}
\label{fig:ska}
\end{figure*}

\section{Results}
\label{sect:results}

Since the diffusion coefficient for dSph galaxies is little-known, different diffusion choices are adopted with $D_0=2.3\times10^{28}\,\rm cm^2s^{-1}$, $\gamma=0.46$ (diffusion choice 1) and $D_0=3\times10^{27}\,\rm cm^2s^{-1}$, $\gamma=0.7$ (diffusion choice 2), respectively \cite{2021JCAP...03..011D}, which are usually used in the context of the Milky Way galaxy. 
It is possible that the diffusion coefficient could be as low as an order of magnitude smaller than that of the Milky Way galaxy, 
so an aggressive condition with  $D_0=3\times10^{26}\,\rm cm^2s^{-1}$, $\gamma=0.3$ (diffusion choice 3) \cite{2019PhRvD.100d3002K,PhysRevD.88.083535} is also adopted.

We place constraints on the PBH abundance using the radio observations of Fornax and Segue I, considering the Hawking radiation from PBHs within the dSphs.
We require that the model-predicted IC/CMB and SSC fluxes do not exceed the upper bound of the $1\sigma$ uncertainty of the radio flux in \cite{2019PhRvD.100d3002K} to set the constraints on the PBH abundance.
The constraints obtained from the Fornax observation are shown in the left panel of Fig.~\ref{fig:mwa}, while the results for the source Segue I are shown in the right panel.
In a relatively narrow mass range ($\lesssim10^{15}\,{\rm g}$), the fraction of dark matter in the form of PBHs is constrained to be {$\lesssim10^{-4}-10^{-5}$} (see the left panels). 
The variation range is due to the choice of the diffusion parameters.
We note that the results are weaker than the existing constraints obtained from the extra-galactic gamma-ray background (EGB) \cite{2010PhRvD..81j4019C} and Voyager 1 data \cite{PhysRevLett.123.251101}. However, by using the SSC approach our work gives the first constraints on the abundance of PBHs in dSphs based on existing radio observations. 
To illustrate the enhancement of the results due to the consideration of the SSC effect, we also {try to} plot the exclusion lines that consider only the IC process in the left panel of Fig.~\ref{fig:mwa}. {However, they} are significantly weaker than the SSC-based results and {cannot appear in the plot} (i.e., the PBH fraction $f_{\rm PBH}$ can not be effectively constrained). 

Future SKA observations will be able to further improve the results of the constraints (in the absence of signal detection).
We here predict the parameters that are expected to be able to be excluded using 100 hours of SKA observations.
The SKA sensitivity curve is taken from \cite{2019PhRvD..99b1302K}.
We first focus on the case considering only SSC. As shown by the red, light blue, and dark blue solid lines of Fig.~\ref{fig:ska}, the expected upper limits from SKA on the PBH abundance would be down to $\lesssim10^{-7}-10^{-8}$. 
Under certain diffusion choices, the constraints at low PBH masses could be improved by a factor of several compared with the existing results in the literature (e.g., the ones derived from the EGB \cite{2010PhRvD..81j4019C} and Voyager 1 data \cite{PhysRevLett.122.041104}).


The above results (i.e., the three SSC lines in Fig.~\ref{fig:ska}) consider only the SSC flux from PBH and ignore the IC/CMB flux to highlight what limits can be obtained by the SSC effect alone. However, since the SSC flux depends on the equilibrium electron density ${\rm d}n/{\rm d}E$ and the number density of synchrotron photons, and the latter also depends on the equilibrium electron density, the SSC flux is in fact more sensitive to the electron spectrum ${\rm d}n/{\rm d}E$ (a second-order dependence) compared to the IC/CMB process. 
This means that when the PBH abundance is low (and thus the electron density due to PBH evaporation is also low), the SSC flux will become very weak. 
For the future SKA observation, if no signal is detected, we can obtain an upper limit on the radio flux quite low (see Fig.~\ref{fig:spectra}), and then even in the waveband close to MWA/GMRT the SSC flux will be lower than the IC/CMB one and no longer dominate. So in the SKA case, it is necessary to consider the IC/CMB flux. 

The constraints on the PBH abundance that can be expected from future SKA observations based on the IC/CMB process are shown as the colored dashed lines in Fig.~\ref{fig:ska} (also considering 3 different diffusion choices). 
It can be seen that in this case stronger constraints are provided by IC/CMB. Although the SKA sensitivity is $\sim$4 orders of magnitude improved compared to MWA/GMRT, the SSC constraints are only 2-order-of-magnitude increased.
Fig.~\ref{fig:ska} also shows that our projected limits are stronger than those ones predicted in Ref.~\cite{2021JCAP...03..011D}, mainly due to the fact that we adopt the dSph Fornax which has a larger diffusion radius (see, e.g., Figs.~\ref{fig:rh} and \ref{fig:mwa}).

We note that the discrepancy between the results for diffusion choices 2 and 3 is small, while the results for the case of diffusion choice 1 (large diffusion coefficient) deviate significantly from the other two. 
This is mainly due to the fact that: the energy loss is dominant when the diffusion is slower, so that the change of the diffusion coefficient has little effect on the results; and the energy loss term has an effect of shifting the IC/CMB spectrum to lower frequencies, which raises the expected flux in the frequency range of SKA (e.g., the solid lines corresponding to d2 and d3 in Fig.~\ref{fig:nb}) and therefore strengthen the constraints. 
For diffusion choice 1, it has crossed the critical point where the energy loss timescale is equal to the diffusion timescale, and diffusion starts to dominate over the energy loss, making the corresponding result deviate from diffusion choices 2 and 3.

\begin{figure}[t!]
\begin{center}
\includegraphics[angle=0,width=0.47\textwidth]{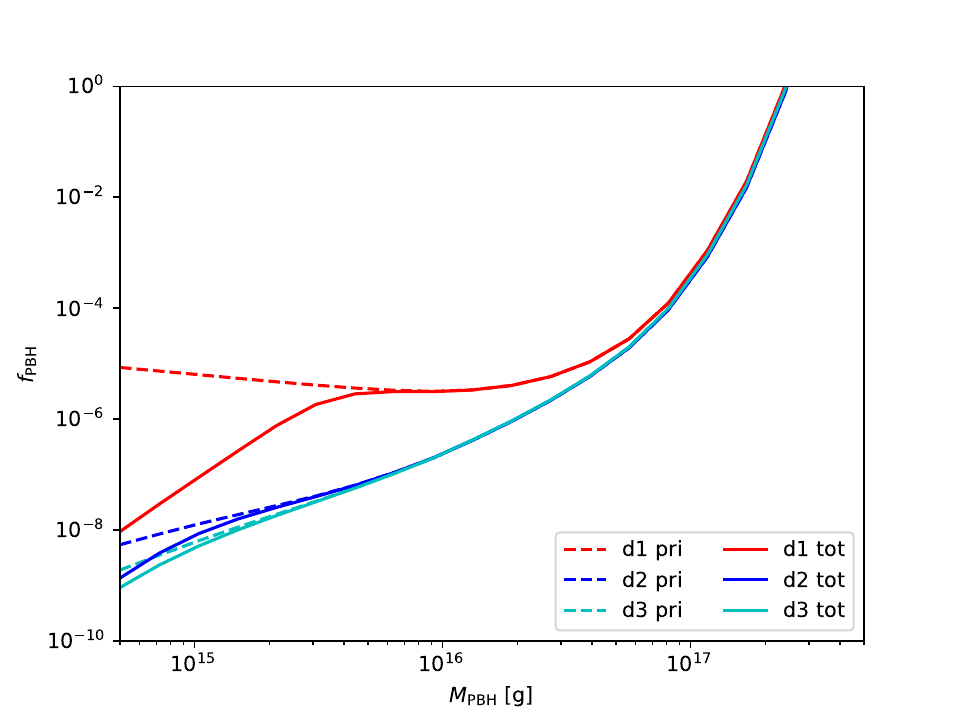}
\end{center}
\caption{The effect of secondary Hawking radiation on the IC/CMB constraints of SKA. The solid (dashed) lines consider (ignore) the secondary Hawking radiation. }
\label{fig:secondary}
\end{figure}

In addition, we note that the results for diffusion choice 1 have a pronounced curvature around {$M_{\rm PBH}\sim3\times10^{15}\,\rm g$}, which is mainly due to the consideration of the secondary particles from the evaporation of the black holes.
As mentioned above, the secondary emission has little influence on the SSC results but is significant for the IC/CMB process at low PBH masses. 
Fig.~\ref{fig:secondary} shows the exclusion lines based on the IC/CMB process with or without the secondary emission of Hawking radiation considered, which demonstrates how the secondary Hawking radiation influences the constraints.

\section{Conclusion}
\label{sect:conclusion}

BH may form in the early universe and could be a DM candidate if its lifetime is larger than the universe's age.
In this paper, we calculate the SSC and IC/CMB fluxes from Hawking radiation electrons of the PBHs in dSphs and constrain the abundance of PBH DM using the MWA and GMRT observational results presented in Ref.~\cite{2019PhRvD.100d3002K}.
We also predict the sensitivity of probing/constraining PBH DM with the future SKA telescope. 
We focus on the PBH mass range of $\sim10^{15}-10^{17}\,\rm g$ because the PBHs with masses smaller than this range have been completely evaporated and those ones larger than this range have very low Hawking radiation flux.

We innovatively consider the SSC process because the SSC peak frequency of the electrons from low-mass PBHs is found to be close to the waveband of the MWA and GMRT radio observations. 
Other improvements/innovations compared to previous works (e.g. \cite{2021JCAP...03..011D}) include that we choose Fornax as the target source for its large diffusion radius.
We also consider the secondary emission from Hawking evaporation.
Considering the secondary emission could give a higher model-predicted radio flux for PBHs at the low mass end and lead to better constraints on the abundance.

These improvements/innovations help us obtain, though weak, the first constraints on the PBH abundance in dSphs at asteroidal mass.
We also obtain projected constraints better than those predicted in previous works (e.g. \cite{2021JCAP...03..011D}) for future SKA observations.

\section{Acknowledgements}
\label{sect:acknowledgements}
We acknowledge the useful discussion with Jian-Kun Huang. This work is supported by the National Key Research and Development Program of China (Grant No. 2022YFF0503304), and the Guangxi Talent Program (“Highland of Innovation Talents”).

\bibliographystyle{apsrev4-1-lyf}
\bibliography{ref}



\end{document}